\documentstyle[11pt,epsfig]{article}
\def\edth{\;\raise1.0pt\hbox{$'$}\hskip-6pt\partial\;}
\def\baredth{\;\overline{\raise1.0pt\hbox{$'$}\hskip-6pt
\partial}\;}
\def\gsim{~\rlap{$>$}{\lower 1.0ex\hbox{$\sim$}}}

\def\be{\begin{equation}}
\def\ee{\end{equation}}
\def\ba{\begin{eqnarray}}
\def\ea{\end{eqnarray}}
\newcommand{\fr}[2]{\frac{#1}{#2}}

\begin{document}

\title{Effects on the two-point correlation function from the coupling of quintessence to dark matter}

\author{Seokcheon Lee$^{\,1,2}$, Guo-Chin Liu$^{\,3,4}$, and Kin-Wang Ng$^{\,1,2,4}$}

\maketitle

$^1${\it Institute of Physics, Academia Sinica,
Taipei, Taiwan 11529, R.O.C.}

$^2${\it Leung Center for Cosmology and Particle Astrophysics, National Taiwan University, Taipei, Taiwan 10617, R.O.C.}

$^3${\it Department of Physics,
Tamkang University, Tamsui, Taipei County, Taiwan 251, R.O.C.}

$^4${\it Institute of Astronomy and Astrophysics,
 Academia Sinica, Taipei, Taiwan 11529, R.O.C.}

\begin{abstract}
We investigate the effects of the nonminimal coupling between the
scalar field dark energy (quintessence) and the dark matter on the
two-point correlation function. It is well known that this
coupling shifts the turnover scale as well as suppresses the
amplitude of the matter power spectrum. However, these effects are
too small to be observed when we limit the coupling strength to be
consistent with observations. Since the coupling of quintessence
to baryons is strongly constrained, species dependent coupling may
arise. This results in a baryon bias that is different from unity.
Thus, we look over the correlation function in this coupled model.
We are able to observe the enhancement of the baryon acoustic oscillation (BAO) peak due to
the increasing bias factor of baryon from this species dependent
coupling. In order to avoid the damping effect of the BAO
signature in the matter power spectrum due to nonlinear
clustering, we consider the coupling effect on the BAO bump in the
linear regime. This provides an alternative method to constrain
the coupling of dark energy to dark matter.

\end{abstract}

Due to the strong constraint on the coupling of the scalar field
dark energy (quintessence) to baryons from the local gravity, we
investigate the effect of the species dependent coupling
\cite{DN,DP} by considering a model in which the quintessence $Q$
is only coupled to the cold dark matter (CDM). We assume a
Yukawa-type coupling, $m_{c} = e^{n{c} Q} m^{*}_{c}$, where
$m_{c}^{*}$ is the bare mass of the CDM~\cite{0307316,LOP}. This
specific choice of coupling requires that the present value of the
scalar field vanishes in order to satisfy $m_{c} = m_{c}^{*}$ at
present. Then we are able to write the general action including
this interaction as \be S = - \int d^4x \sqrt{-g} \Biggl\{
\frac{\bar{M}^2}{2} \Bigl[ -R +
\partial^{\mu} Q
\partial_{\mu} Q \Bigr] + V(Q) - {\cal L}_{c}  - {\cal L}_{r}
- {\cal L}_{b} \Biggr\}, \label{S} \ee where $\bar{M} = 1/
\sqrt{8\pi G}$ is the reduced Planck mass, $V(Q)$ is the potential
of $Q$, and ${\cal L}_{r}$ and ${\cal L}_{i} = - m_{i}
\delta(\vec{x} - \vec{x}_{i}(t)) \sqrt{g_{\mu\nu}
\dot{x}_{i}^{\mu} \dot{x}_{i}^{\nu} / g}$ ($i = c, b$) denote the
Lagrangian of radiation, CDM, and baryons respectively. We adopt
$V(Q) = V_{0} \exp (\lambda Q^2/2)$ with $\lambda
= 5$ in the following \cite{LOP,LLN}. However, the main
conclusions are independent of the form of the scalar field
potential (see below). Due to the coupling, the scalings of the CDM and the
quintessence energy densities are changed respectively
to~\cite{LLN} \ba \rho_c (a) &=& \rho_{c}^{0} a^{-3 + \epsilon} \,
\hspace{0.1in} {\rm where} \hspace{0.1in} \epsilon \ln(a) = n_{c}
\Bigl[ Q(a) - Q(1) \Bigr] \, ,  \label{rhoc} \\ \rho_{Q}' &=& -3
{\cal H} (1 + \omega_{Q}^{\rm{eff}}) \rho_{Q} \,\, \hspace{0.1in}
{\rm where} \hspace{0.1in} \omega_{Q}^{\rm{eff}} = \omega_{Q} +
\fr{n_{c}}{{\cal H}} \fr{\rho_{c}}{\rho_{Q}} Q' \, , \label{drhoQ}
\ea  where ${\cal H} \equiv (da/d \eta)/a$, $\omega_{Q}$ is the
equation of state (eos) of the quintessential dark energy (DE),
$\rho_{c}^{0}$ denotes the present value of the CDM energy
density, the present value of scale factor $a_{0} = 1$, and primes
mean the differentiation with respect to the conformal time
$\eta$. Generally, the sign of $\epsilon$ depends on both the
model and the form of the coupling. The linear perturbation
equations for the CDM and the scalar field $Q$ in the synchronous
gauge are \cite{Luca1}
\ba \delta_{c}' &=& - \theta_{c} - \fr{1}{2} h' + n_c \delta Q' \, , \label{deltac} \\
\theta_{c}' &=& -{\cal H} \theta_{c} + n_{c} \Bigl( k^2 \delta Q -
Q' \theta_{c} \Bigr) \, , \label{thetac} \\ \delta Q'' &+& 2 {\cal
H} \delta Q' + k^2 \delta Q + \fr{a^2}{\bar{M}^2} \fr{\partial^2
V}{\partial Q^2} \delta Q = -\fr{1}{2} h' Q' - \fr{a^2}{\bar{M}^2}
\rho_{c} n_{c} \delta_{c} \, , \label{deltaQ} \ea where $k$ is the
wave-number, $h$ is the metric perturbation,
$\delta_c=\delta\rho_c/\rho_c$, and $\theta_c$ is the gradient of
the CDM velocity flow. Also from the perturbed Einstein equations,
we obtain \be \fr{1}{2} (h'' + {\cal H} h') = - \fr{a^2}{2
\bar{M}^2} \Bigl[ 2 \delta \rho_r + \delta \rho_{c} + (1 +
3c_{b}^2) \delta \rho_b + \fr{4 \bar{M}^2}{a^2} Q' \delta Q' - 2
\fr{\partial V}{\partial Q} \delta Q \Bigr] \, , \label{heq} \ee
where $c_b$ is the sound speed of baryons. Note that we will adopt the adiabatic initial conditions and
thus $k^2\delta_c$ term is absent in Eq.~(\ref{thetac}).

The coupling strength $n_{c}$ is commonly constrained through the comparison with
the observed cosmic microwave background anisotropy and matter power spectrum.
In Ref.~\cite{LLN}, we found $n_{c}\le 0.01$. Even though the actual value of the upper limit
depends on the form of the quintessence potential and that of the coupling,
the obtained limits for other potentials and couplings are of the same order
as shown in Ref.~\cite{bean}. Furthermore, Eqs.~(\ref{deltac}) and (\ref{thetac}),
which describe the evolutions of the CDM density and velocity field respectively,
imply that the influence of the quintessence field are dwarfed by the background evolution, ${\cal H}$.
Thus, as long as we have the late-time dominated quintessence model,
the evolution behaviors of Eqs.~(\ref{deltac}) and (\ref{thetac}) are
quite similar and almost independent of the form of potential.
However, we have also checked the early dark energy model, for example, with
the potential given by $V(Q)=V_0 \cosh(\lambda Q)$, in which the dark energy component
is not negligible at early times~\cite{wolung}.
In this model, ${\cal H}$ is quite different from the quintessence model with
the exponential potential and produce a quite different behavior of $\delta_c$.
Here we will concentrate on the late quintessence model.

Even though we use the
full set of the above equations for our calibration, we
investigate the effects of coupling on the matter power spectrum
with some approximation in what follows. It is well known that
averaging out the small and oscillatory $\delta Q''$ and $\delta
Q'$ is a good approximation for all scales~\cite{Koivisto}. From
this fact, we obtain the approximate expression of $\delta Q$ from
Eq.~(\ref{deltaQ}) \be \delta Q \simeq - \fr{\fr{a^2}{\bar{M}^2}
n_c \rho_c \delta_c + \fr{1}{2} Q' h'}{\fr{a^2}{\bar{M}^2} m_{Q}^2
+ k^2} \, , \label{deltaQ2} \ee where $m_{Q}^2 \equiv \partial^2 V
/\partial Q^2$. We obtain the evolution equation of $\delta_c$
from Eqs.~(\ref{deltac}) and (\ref{thetac}) by using Eq.
(\ref{deltaQ2}), \be \delta_{c}'' + {\cal H} \Bigl[ 1 + n_{c}
\sqrt{3(1+\omega_{Q}) \Omega_{Q}} \, \Bigr] \delta_{c}' -
\fr{3}{2} {\cal H}^2 \Omega_{c} \Bigl( 1 + 2n_c^2 \Bigr)
\delta_{c} \simeq 0 \, , \label{dddeltac} \ee where $\Omega_Q$ is
the quintessence energy density relative to the critical density
and we use $Q' = {\cal H} \sqrt{3(1+\omega_{Q}) \Omega_{Q}}$~
\cite{LOP}. The evolution of the linear perturbation of the baryon
$\delta_b$ which is not coupled to the scalar field is the same as
the above equation~(\ref{dddeltac}) except that now the coupling
terms are absent: \be \delta_{b}'' + {\cal H} \delta_{b}' -
\fr{3}{2} {\cal H}^2 \Omega_{c} \delta_{c} \simeq 0 \, .
\label{dddeltab} \ee
It is convenient to rewrite the above equations~(\ref{dddeltac})
and (\ref{dddeltab}) in terms of $x = \ln a$ as \ba && \fr{d^2
\delta_{c}}{dx^2} + \Biggl[ \fr{1}{2} -\fr{3}{2} \omega_{Q}
\Omega_{Q}  + n_{c} \sqrt{3(1+\omega_{Q}) \Omega_{Q}} \, \Biggr]
\fr{d \delta_{c}}{dx} \nonumber \\ && - \fr{3}{2} ( 1 + 2n_c^2 )
\Omega_{c} \delta_{c} \simeq 0 \, , \label{dddeltacx} \\ &&
\fr{d^2 \delta_{b}}{dx^2} + \Biggl(\fr{1}{2} -\fr{3}{2} \omega_{Q}
\Omega_{Q} \, \Biggr) \fr{d \delta_{b}}{dx} - \fr{3}{2} \fr{1}{b}
\Omega_{c} \delta_{b} \simeq 0 \, , \label{dddeltabx} \\ && {\rm
where} \,\, \Omega_{Q} \simeq 1 - \Omega_{c} \simeq \Biggl[ 1 +
\fr{\Omega_{c}^{0}}{\Omega_{Q}^{0}} \fr{e^{(-3 + \epsilon
)x}}{e^{-3 \int_{0}^{x} [1 + \omega_{Q}(x')] dx'}} \Biggr]^{-1}
\,\, . \label{OmegaQ2} \ea In Eq.~(\ref{dddeltabx}), we define a
baryon bias factor $b$ by $\delta_{b} \equiv b \delta_{c}$ with a
small time variation ({\it i.e.} we ignore $db/dx$ and $d^{2}
b/dx^2$ terms). If we further use the assumption that the linear
growth of the CDM is given by $\delta_{c} \propto e^{m x}$ with $m
< 1$, then we will obtain the analytic form of $b$ \be b = \Biggl[
1 + 2 n_{c}^2 - \fr{2 n_{c} m \sqrt{3 (1 + \omega_{Q})
\Omega_{Q}}}{3 \Omega_{c}} \Biggr]^{-1} \, . \label{b} \ee
\begin{center}
\begin{figure}
\vspace{1.5cm} \centerline{ \psfig{file=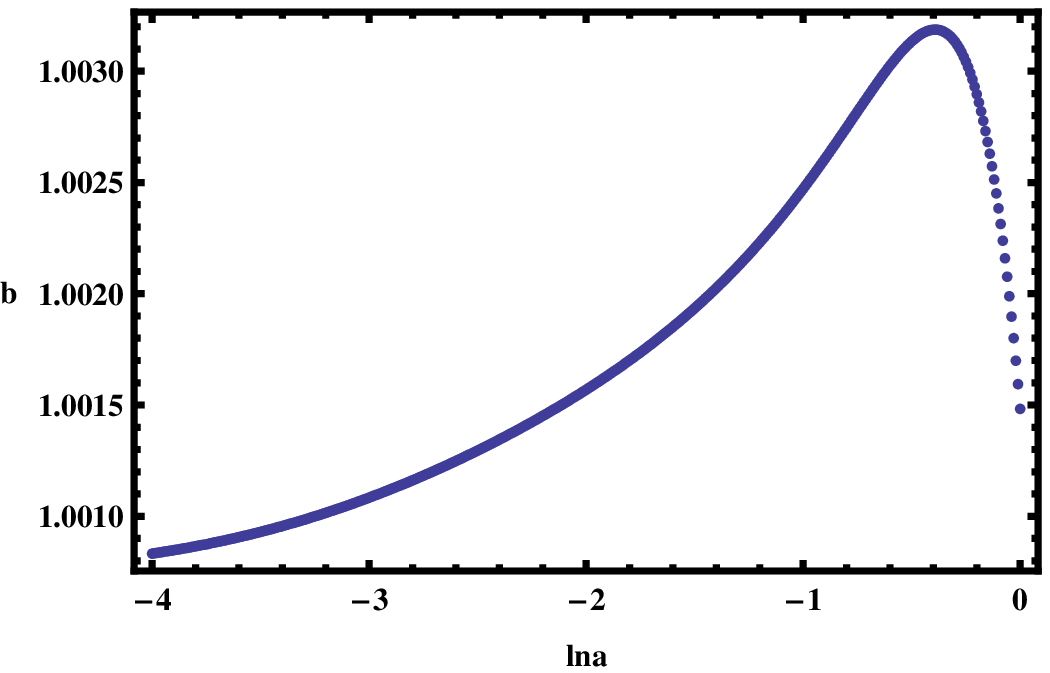, width=6.5cm}
\psfig{file=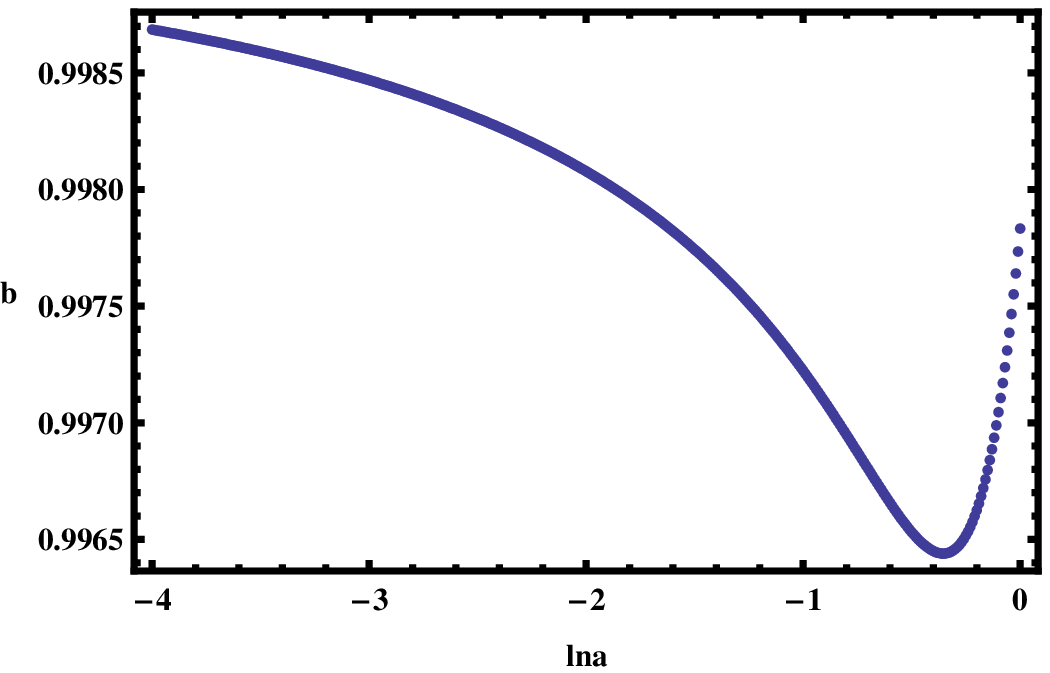, width=6.5cm} } \vspace{-0.5cm} \caption{
Evolutions of the baryon bias factor $b$ for the different
couplings when $n_{c} = 0.01$ (left panel) and $n_{c} = -0.01$
(right panel) } \label{fig1}
\end{figure}
\end{center}
Here we use the ansatz for the coupling between CDM and DE as
$e^{n_{c}Q}$. In our model, the scalar field evolves from $-5$ to
$0$ during $10^{-10} \leq a \leq 1$. We put a limit on the
magnitude of the coupling constant $|n_{c}| \leq 0.01$, where the
choice of the sign of $n_{c}$ is still arbitrary. The effective
mass of the CDM varies at most around $5$ \%. $m_{c}$ ({\it
equally}, $\rho_{c}$) increases (decreases) as it evolves to the
present for the positive (negative) $n_{c}$. Thus, the evolutions
of the background quantities are slightly changed dependent on the
sign of $n_{c}$. The effect of the coupling on the $b$ also
depends on the sign of $n_{c}$ as given in Eq.~(\ref{b}), which is used
to estimate $b$ as shown in Fig.~\ref{fig1}. In the left panel of Fig. \ref{fig1}, we show
the evolution of $b$ when $n_{c} = + 0.01$. For the positive
$n_{c}$, the last term in Eq.~(\ref{b}) is negative and its
magnitude is bigger than $2n_{c}^2$ for the given model. Thus, $b
\geq 1$. The evolution of $b$ for $n_{c} = -0.01$ is depicted in
the right panel of Fig.~\ref{fig1}. In this case, the last term in
Eq.~(\ref{b}) is positive and $b$ is always smaller than $1$.
Thus, we are able to constrain not only the magnitude but also the
form of the coupling between CDM and DE from accurate observations
of the baryon power spectrum. A similar but slightly different
conclusion was drawn in Ref. \cite{Luca}; however, their
conclusion is only true for the tracking region solutions.
\begin{center}
\begin{figure}
\vspace{1.5cm} \centerline{ \psfig{file=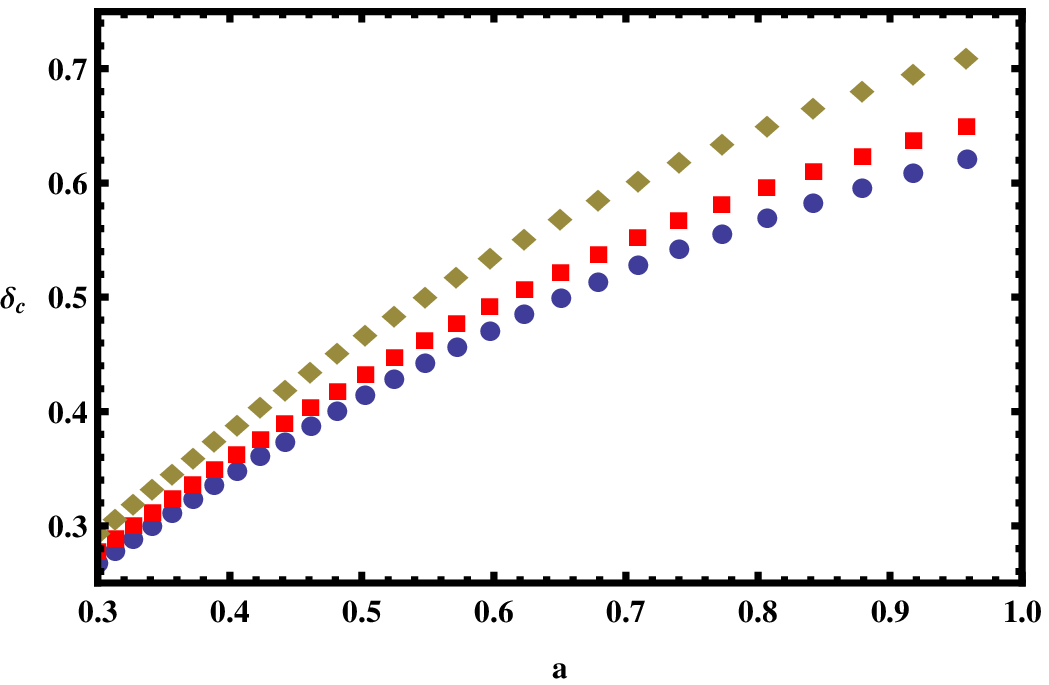,
width=6.5cm}\psfig{file=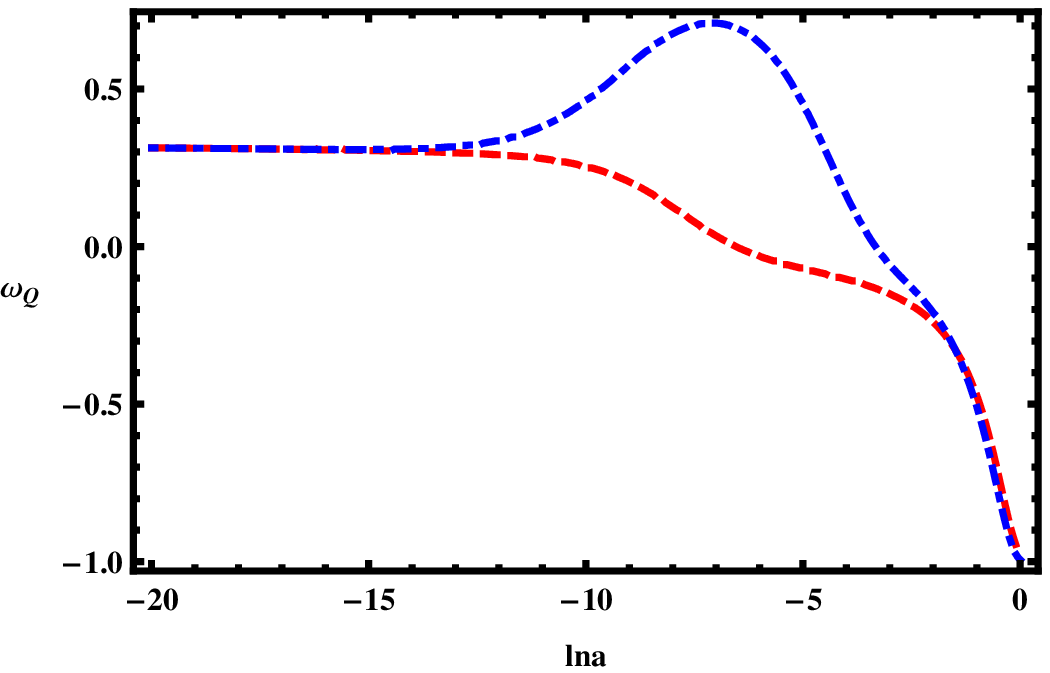, width=6.5cm} } \vspace{-0.5cm}
\caption{ a) Evolutions of $\delta_{c}$ in the $\Lambda$CDM, QCDM,
and cQCDM models (from top to bottom). b) Evolutions of the
equation of state of Q field, $\omega_{Q}$, for the coupled and
noncoupled cases (from top to bottom).} \label{fig2}
\end{figure}
\end{center}

First, we study the CDM density fluctuation $\delta_c$ given in
Eq. (\ref{dddeltacx}) for different cases. We denote respectively
the cosmological model including CDM component with the
cosmological constant as $\Lambda$CDM, with the non-coupled Q
field ($n_{c} = 0$) as QCDM, and with the coupled Q field as
cQCDM. We also denote $\delta_{c}$ for each model as
$\delta_{c}^{\Lambda CDM}$, $\delta_{c}^{QCDM}$, and
$\delta_{c}^{cQCDM}$. $\delta_{c}^{QCDM}$ has the same evolution
equations as those of the $\Lambda$CDM model except the difference
in $\Omega_{Q}$ given in Eq.~(\ref{OmegaQ2}). $\omega_{Q}$ in the
QCDM model changes from $1/3$ in the radiation-dominated epoch
(the so-called ``early tracking region'') to around $-1$ at
present. Thus, $\omega_{Q} > \omega_{\Lambda} = -1$ during the
entire epoch. This causes the suppression of $\delta_{c}$ in QCDM
model compared to $\Lambda$CDM model when we use the same present
values of the cosmological parameters. We illustrate this in the
left panel of Fig. \ref{fig2}. The diamond and the rectangular
points correspond to $\delta_{c}^{\Lambda CDM}$ and
$\delta_{c}^{QCDM}$, respectively. We also compare
$\delta_{c}^{QCDM}$ and $\delta_{c}^{cQCDM}$. When the CDM is
coupled to Q field, the scaling of $\rho_{m}$ is changed as given
in Eq. (\ref{rhoc}). Also $\omega_{Q}$ is increased during the
matter-domination epoch in the cQCDM model as shown in the right
panel of Fig.~\ref{fig2}. The dot-dashed and the dashed lines
depict $\omega_{Q}$ when $n_{c} = 0.01$ and $0$, respectively.
Thus, this causes slightly further suppression of $\delta_{c}$ in
the cQCDM model. However, if we want to compare
$\delta_{c}^{cQCDM}$ with $\delta_{c}^{QCDM}$ at the relevant
sub-horizon scale, then we should constrain their evolutions at
late times $a \geq 0.1$ ({\it equally}, $x \geq -2.3$). But then
$\omega_{Q}$s of the two models are almost identical as shown in
the right panel of Fig. \ref{fig2} and the discrepancy between
$\delta_{c}^{cQCDM}$ and $\delta_{c}^{QCDM}$ becomes negligible.
If we use the definition of the baryon bias factor $b =
\fr{\delta_{b}}{\delta_{c}}$, then we would obtain that $b_{cQCDM}
> b_{QCDM} \simeq b_{\Lambda CDM} \simeq 1$ for the same present
values of the cosmological parameters.

\begin{center}
\begin{figure}
\vspace{1.5cm}
\centerline{
\psfig{file=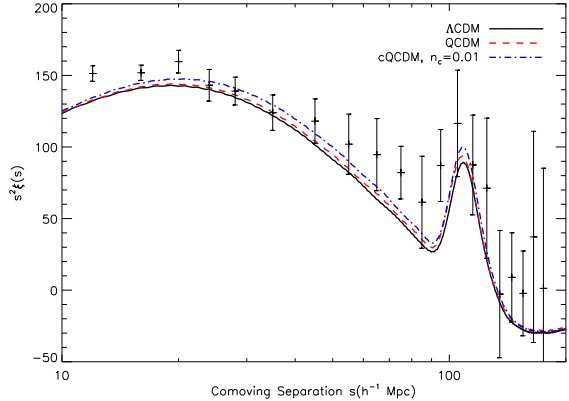, width=8cm} }
\vspace{-0.5cm}
\caption{Matter correlation functions for $\Lambda$CDM, QCDM, and cQCDM models.} \label{fig3}
\end{figure}
\end{center}

Now we consider the previously mentioned differences in the
two-point correlation functions in different models. Instead of using the above approximations,
we will run the numerical evolution of the full system of equations.
The coupling of the quintessence to the dark matter modifies both the turnover
scale and the amplitude of the matter power spectrum. However, the
shift in the turnover scale and the suppression in the amplitude
of the matter power spectrum (defined by $P(k) = \langle |
\delta_{k} |^2 \rangle \equiv (2 \pi^2)/k^3 \Delta^2(k)$) due to
this coupling are too small when we limit the coupling strength to
be consistent with observations. Thus, this gives us the
motivation to probe the coupling effects on the baryon acoustic
peak in the correlation function, $\xi(s) = \int \Delta^2(k)
J_{0}(ks) d \ln k$, where $J_{0}$ is the spherical Bessel
function. In order to avoid the damping effect of the BAO
signature in the matter power spectrum due to nonlinear
clustering, we put the limitation of $k \le 0.2 h \, {\rm
Mpc}^{-1}$ in our correlation function calculation. We show the
correlation function times the comoving separation square ($s^2$)
in Fig.~\ref{fig3}. The solid, dashed, and dotted lines correspond
to the $\Lambda$CDM, QCDM, and cQCDM models, respectively. The
cosmological parameters that we use in this figure are $H_{0} =
71\,{\rm km/sec/Mpc}$, $\Omega_{b} = 0.047$, $\Omega_{c} = 0.211$,
$\Omega_{Q} = 0.742$, and the galaxy bias factor $b_{gal} = 1.9$.
We also normalize the matter power spectrum to $\sigma_{8} =
0.788$ to be consistent with the WMAP and the Luminous Red
Galaxies (LRG) observations~\cite{tegmark}. Note that the influence of DE on the
present value of $\sigma_{8} = \sqrt{\int_{0}^{k} W^2(kR_{8})
\Delta^2(k) d \ln k}$, where $W(x) = 3(\sin x/x^3 - \cos x /x^2)$
is the Fourier transform of a top-hat window function and $R_{8} =
8 h^{-1} {\rm Mpc}$, is well studied
\cite{WS,0107525,0206507,0307346}.

In Fig.~\ref{fig3}, we are able to clearly see the differences of
the correlation functions for the different models. The first peak
in the correlation function corresponds to the turnover scale
which is related to the scale factor $a_{eq} \sim 1/3500$ when the
radiation and the matter densities are equal. The $\Lambda$CDM and
QCDM models have the same $a_{eq}$ for the same set of
cosmological parameters while their comoving distances of the
matter and radiation equality, $s_{eq}$, are $19.17$ ${\rm Mpc}/h$
and $19.60$ ${\rm Mpc}/h$, respectively. The discrepancy comes
from the fact that the comoving distance to the $a_{eq}$ is
$\int_{0}^{t_{eq}} \fr{dt'}{a(t')} = \int_{z_{eq}}^{\infty}
\fr{dz'}{H(z')}$ and the two models have slightly different
$H(z)$. Also, the location of $s_{eq}$ in the cQCDM model is
different from that in the QCDM model due to the change in the
scaling of the CDM density. In our model, $\epsilon
> 0$ and it causes the delay of the radiation and matter equality
epoch ($a_{eq}^{cQCDM} > a_{eq}^{QCDM}$). Thus, the comoving
distance $s_{eq}$ is shifted to about $20.87$ ${\rm Mpc}/h$ in the
coupled case.

We have already shown for the relevant $k$-modes, $\delta_{c}^{\Lambda CDM}
> \delta_{c}^{QCDM} \simeq \delta_{c}^{cQCDM}$ for the same present values
of the cosmological parameters. Thus, if we normalize the power spectra of
the latter two models on cluster scales $\sigma_{8}$, they acquire a larger
amplitude of primordial fluctuations compared to the $\Lambda$CDM model.
The BAO bump in the correlation function of both QCDM and cQCDM
models are larger than that of the $\Lambda$CDM model. However,
the reason for the enhancements of the BAO bumps in both models
are different. The enhancement in the QCDM model compared to the
$\Lambda$CDM one is due to the choice of the larger amplitude of
primordial fluctuations. The enhancement in the cQCDM model is due
to the coupling between Q and CDM. We observe this effect in
Fig.~\ref{fig3}. If we compare the BAO peak of cQCDM with that of
$\Lambda$CDM or QCDM, then we observe that it is enhanced in the
cQCDM model. This is consistent with our early explanation that the baryon bias factor
in the cQCDM model is enhanced as given in Eq.~(\ref{b}). The overall amplitudes of all three models in Fig.~\ref{fig3} are
smaller than those given in Ref.~\cite{0501171}. This is due to
the different choices of the galaxy bias factor. We use the LRG galaxy bias factor $b_{gal} =1.9$~\cite{tegmark} whereas the authors
in Ref.~\cite{0501171} claim that they use the scale dependence bias factor
which is however not given therein. However, both the shapes of the correlation functions and
the amplitudes of the BAO bumps of the QCDM and cQCDM models fit
to the data points better than the $\Lambda$CDM model. As we
mentioned above, if we choose a negative $n_{c}$, then the
amplitude of the BAO bump in the cQCDM model is decreased due to
anti-biasing of baryons with $b < 1$. This effect mimics the
nonlinear effect \cite{07042783}. It will suppress the amplitude
and shift the location of the BAO bump, and thus making harder for
us to fit the data. However, we limit the calculation in the
linear regime and this effect is irrelevant for our consideration.
In the Figure~3 of Ref.~\cite{0501171}, it is claimed that
$\Omega_{c}h^2 = 0.13$ shows the better fit to the data compared
to the case using $\Omega_{c}h^2 = 0.12$. However, the amplitude
of the BAO bump with $\Omega_{c}h^2 = 0.13$ shows the bigger
discrepancy with the data than that with $\Omega_{c}h^2 = 0.12$.
We are able to give a better direction for fitting both the shape and
the amplitude of the correlation function in the QCDM and cQCDM models.
We have done a simple $\chi^2$ analysis and found that the
$\chi^2$ values are $21.6$, $21.7$, and $23.4$ for the $\Lambda$CDM,
QCDM, and cQCDM models, respectively.
Although the improvement has low statistical significance,
the species-dependent coupling effects to the correlation function
in the coupled quintessence model may be potentially important
and are being further studied. Note that there is a slight shift in the location of the BAO peak
in the QCDM model. This is due to the change in $H(z)$ in the QCDM
model compared to the $\Lambda$CDM model. This affects the sound
horizon that is given by $l_s=\int_{z_{dec}}^{\infty}
\fr{c_{s}}{H(z)} dz$, where the sound speed of the baryon-photon
fluid $c_{s} = 1/\sqrt{3(1+ 3\rho_{b}/4\rho_{r})}$ is the same for
every model. Hence, for each model we have a different $H(z)$ and
$l_s$ will vary. However, the difference is quite small.

\section*{Acknowledgments}
This work was supported in part by the National Science Council, Taiwan, ROC under the Grants NSC NSC 97-2112-M-032-007-MY3 (GCL), 98-2112-M-001-009-MY3 (KWN), and the National Center for Theoretical Sciences, Taiwan, ROC.


\begin{thebibliography}{99}

\bibitem{DN}T.~Damour and K.~Nordtvedt,
Phys.\ Rev.\ Lett.\  {\bf 70}, 2217 (1993); T.~Damour and K.~Nordtvedt,
Phys.\ Rev.\ D {\bf 48}, 3436 (1993).

\bibitem{DP} T. Damour and A. M. Polyakov, Nucl.\ Phys.\ B {\bf 423}, 532 (1994)
[arXiv:hep-th/9401069]; Gen.\ Rel.\ Grav {\bf 26}, 1171 (1994) [arXiv:gr-qc/9411069].

\bibitem{0307316} G.~R.~Farrar and P.~J.~E.~Peebles, Astrophys.\ J.\ {\bf 604}, 1 (2004) [arXiv:astro-ph/0307316].

\bibitem{LOP} S.~Lee, K.~A.~Olive, and M.~Pospelov, Phys.\ Rev.\ D {\bf 70}, 083503 (2004) [arXiv:astro-ph/0406039]. 

\bibitem{LLN} S.~Lee, G.-C.~Liu, and K.-W.~Ng, Phys.\ Rev.\ D {\bf 73}, 083516 (2006) [arXiv:astro-ph/0601333]. 

\bibitem{Luca1} L.~Amendola, Phys.\ Rev.\ D {\bf 69}, 103524 (2004) [arXiv:astro-ph/0311175]. 

\bibitem{bean} R.~Bean, E.~E.~Flanagan, I.~Laszlo, and M.~Trodden, Phys.\ Rev.\ D {\bf 78}, 123514 (2008) [arXiv:0808.1105]. 
; G.~L.~Vacca, J.~R.~Kristiansen, L.~P.~L.~Colombo, R.~Mainini, and S.~A.~Bonometto, JCAP 0904:007 (2009) [arXiv:0902.2711]. 

\bibitem{wolung} W. Lee and K.-W. Ng, Phys.\ Rev.\ D {\bf 67}, 107302 (2003) [arXiv:astro-ph/0209093]. 
; R.~R.~Caldwell, M.~Doran, C.~M.~Mueller, G.~Schaefer, and C.~Wetterich, Astrophys.\ J.\ {\bf 591}, L75 (2003) [arXiv:astro-ph/0302505]. 

\bibitem{Koivisto} T.~Koivisto, Phys.\ Rev.\ D {\bf 72}, 043516 (2005) [arXiv:astro-ph/0504571].

\bibitem{Luca} L.~Amendola and D.~Tocchini-Valentini, Phys.\ Rev.\ D {\bf 66}, 043528 (2002) [arXiv:astro-ph/0111535]. 

\bibitem{tegmark} M.~Tegmark {\it et\ al.\,}, Phys.\ Rev.\ D {\bf 74}, 123507 (2006) [arXiv:astro-ph/0608632]. 

\bibitem{WS} L.~Wang and P.~J.~Steinhardt, Astrophys.\ J. {\bf 508}, 483 (1998) [arXiv:astro-ph/9804015].

\bibitem{0107525} M.~Doran, J.-M.~Schwindt, and C.~Wetterich, Phys.\ Rev.\ D {\bf 64}, 123520 (2001) [arXiv:astro-ph/0107525]. 

\bibitem{0206507} M.~Bartelmann, F.~Perrotta, and C.~Baccigalupi, Astron.\ Astrophys.\ {\bf 396}, 21 (2002) [arXiv:astro-ph/0206507].

\bibitem{0307346} M.~Kunz, P.-S.~Corasaniti, D.~Parkinson, and E.~J.~Copeland,
Phys.\ Rev.\ D {\bf 70}, 041301 (2004) [arXiv:astro-ph/0307346].

\bibitem{0501171} D.~J.~Eisenstein {\it et\ al.\,}, Astrophys.\ J. {\bf 633}, 560 (2005) [arXiv:astro-ph/0501171].

\bibitem{07042783} M.~Crocce and R.~Scoccimarro, Phys.\ Rev.\ D {\bf 77}, 023533 (2008) [arXiv:0704.2783].









\end{thebibliography}
\end{document}